\documentclass[aps,prb,twocolumn,showpacs,superscriptaddress,groupedaddress]{revtex4-1}  
\usepackage{graphicx}  
\usepackage{dcolumn}   
\usepackage{bm}        
\usepackage{amssymb}   
\usepackage{siunitx}
\usepackage{amsmath}
\usepackage{amssymb,amstext}
\usepackage{upgreek}
\usepackage{braket}
\usepackage{ marvosym }
	
\begin{document}

\widetext

\hspace{5.2in} 

\title{Narrowing of the Overhauser field distribution by feedback-enhanced dynamic nuclear polarization}
\author{S. Tenberg}
\altaffiliation[Now at: ]{Centre for Quantum Computation and Communication Technology, School of Electrical Engineering
and Telecommunications, University of New South Wales, Sydney, New South Wales 2052, Australia}
\author{R.P.G. McNeil}
\affiliation{JARA-Institute for Quantum Information, RWTH Aachen University, D-52074 Aachen, Germany}
\author{S. Rubbert}
\altaffiliation[Now at: ]{Kavli Institute of NanoScience, Delft University of Technology, Delft, The Netherlands}
\author{H. Bluhm}
\affiliation{JARA-Institute for Quantum Information, RWTH Aachen University, D-52074 Aachen, Germany}
\date{\today}

\begin{abstract}
In many electron spin qubit systems coherent control is impaired by the fluctuating nuclear spin bath of the host material. Previous experiments have shown dynamic nuclear polarization with feedback to significantly prolong the inhomogeneous dephasing time $T_2^*$ by narrowing the distribution of nuclear Overhauser field fluctuations.  We present a model relating the achievable narrowing of the Overhauser field to both the pump rate and the noise magnitude and find reasonable agreement with experimental data. It shows that former experiments on gated GaAs quantum dots were limited by the pump rate of the pumping mechanism used. Here we propose an alternative feedback scheme using electron dipole spin resonance. Sequentially applying two ac electric fields with frequencies slightly detuned from the desired Larmor frequency results in a pump curve with a stable fixed point. Our model predicts that $T_2^*$ values on the order of microseconds can be achieved. 
\end{abstract}

\maketitle

One limiting factor of spin qubits based on semiconductor quantum dots is the fluctuation of the effective magnetic field arising from nuclear spins. The interaction of these nuclear spins in the host material with the electron spin in the quantum dot induces decoherence. Several techniques have been developed to reduce decoherence.\cite{Danon2009a, Rudner2007, Xu2009, Klauser2006, Bluhm2010,Ladd2010,Hogele2012,Shulman2014a} Narrowing the distribution of the nuclear fluctuations was shown to prolong the inhomogeneously broadened dephasing time for gate defined GaAs quantum dots\cite{Bluhm2010} and for self-assembled InAs quantum dots.\cite{Hogele2012, Xu2009}
To reduce fluctuations the nuclear field is regulated by dynamic nuclear polarization (DNP). DNP controls the Overhauser field by transferring the electron spin angular momentum to the nuclear spins via the hyperfine interaction. If the effectiveness of the polarization step depends on the current state of the nuclear spin bath, the qubit can be used as a closed feedback loop. 

In gate defined GaAs double quantum dots, a roughly tenfold enhancement of $T_2^*$ has been achieved by Landau-Zener-sweep driven  electron-nuclear spin flip-flops that were preceded by a free evolution of the qubit to create a feedback effect.\cite{Bluhm2010} In self-assembled InAs quantum dots a DNP feedback effect was also observed as the quantum dot resonance exhibited locking to frequency \cite{Hogele2012, Latta2009, Xu2009} or repetition rate \cite{Greilich2007} of the incident laser. Driving GaAs gate defined quantum dots with electron dipole spin resonance (EDSR) showed locking to the spin-resonance condition as well.\cite{Vink2009}

The narrowing effect has been examined theoretically in various studies but these were focussed on microscopic models and optical manipulation.\cite{Yang2013, Barnes2012, Economou2014, Urbaszek2013} A phenomenological rate-equation model to describe the effectiveness of DNP in reducing the Overhauser field variance was derived by Vink \textit{et al.} \cite{Vink2009} and by Latta \textit{et al}.\cite{Latta2009} The model relates the maximum nuclear field achievable by polarization to the nuclear field variance by treating the polarization as a random variable influenced by DNP and relaxation proportional to the current value. This model was further refined by Yang \textit{et al}.\cite{Yang2012} However, the dominant relaxation process in gated dots is spin diffusion of the local polarization to its surroundings, which is not a simple relaxation process.\cite{Reilly2008,Sallen2014} As a result, the maximum field is determined by the slow long range nuclear dynamics, while the variance of the fluctuations also depends on short range fluctuations. Overall, diffusive dynamics have received little attention, but are of particular interest for gated dots. 

In this paper we present a model relating the achievable narrowing of the nuclear field distribution to the pump rate and the noise magnitude by explicitly considering the different relaxation rates for polarization modes with different wave length. To this end we solve the spin diffusion equation including Langevin dynamics. The results are in reasonable agreement with experimental data and show that the variance of nuclear fluctuations depends directly on the pump rate via the feedback gain. 
Previous experiments were limited by the pump rate, for example in gate defined GaAs dots the pump rate was set by slow Landau-Zener sweeps at the avoided crossing of the singlet and $T_+$ state ('ST$_+$ scheme').\cite{Bluhm2010} Thus, to overcome the limitation of the ST$_+$ scheme, we propose a different polarization scheme based on the work of Laird \textit{et al},\cite{Laird2007} the 'EDSR scheme'. Here an ac electric field is used to flip the spin of the electron in the dot via the hyperfine interaction. Simultaneously a nuclear spin is flipped. High spin flip rates 
can be achieved as no time is lost in Landau-Zener sweeps and bidirectional feedback is possible, thereby eliminating the feedback-free reverse pumping necessary for the ST$_+$ scheme. These high spin flip rates promise further narrowing of the nuclear field distribution. Our calculations indicate that dephasing times of up to $T_2^*=6\,\upmu$s are possible with the EDSR scheme. 

The remainder of the paper is organized as follows. In section \ref{sec:model} we first briefly review the nuclear dynamics and then present the spin diffusion model relating pump rate and noise. Section \ref{sec:comp} compares the results of the model to experimental data. EDSR and the new EDSR scheme for a single quantum dot are described and the achievable narrowing is evaluated in section \ref{sec:polscheme}. Section \ref{sec:dd} shows how the EDSR scheme can be applied to a double dot system with and without exchange coupling. A summary and discussion of results is given in section \ref{sec:sum}. 

\section{Nuclear dynamics and spin diffusion model \label{sec:model}} 

\subsection{Nuclear spins}

\begin{figure}[htp]
  \centering
      \includegraphics[width=0.4\textwidth]{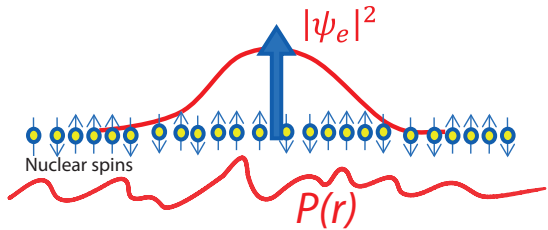}
  \caption[Illustration of the interaction between nuclear spins and the electron]{Illustration of the interaction between nuclear spins (small arrows) and the electron spin (big arrow) in the dot. If the wave function of the nuclei and the electron wave function overlap, a finite probability of a simultaneous electron and nuclear spin flip occurs due to hyperfine contact interaction. The spatially varying polarization of the nuclei can be described by the function $P(r)$. This polarization influences the electron's Larmor frequency.}
  \label{fig:nucspins}
\end{figure}

The nuclear spins of the host material and the electrons in the dot interact via the Fermi contact hyperfine interaction   
described by
\begin{equation}
H_{\rm{hf}}= \nu_0 \sum_i \mathcal{A}_{\alpha(i)} |\psi_i(\mathbf{R}_i)|^2\mathbf{S}_e\cdot\mathbf{I}_i
\end{equation}
 as illustrated in figure \ref{fig:nucspins}. Here, $\mathcal{A}_{\alpha(i)}$ is the microscopic hyperfine coupling for the different nuclear species $\alpha(i)$, which is weighted by the local electron density $|\psi_i(\mathbf{R}_i)|^2$, and $\nu_0$ is the volume of the primitive unit cell containing two nuclei (in III-V materials). 
$\mathbf{S}_e$ is the electron spin and $\mathbf{I}_i$ the angular momentum operator of the \textit{i}th nuclear spin. 

In consideration of the large number of nuclear spins interacting significantly with the electron (typically $10^6$), we describe their effect in terms of a classical effective magnetic field, the Overhauser field,
\begin{equation}
\mathbf{B}_{\rm{nuc}}=\frac{\nu_0 A}{g_{\rm e}^*\mu_{\rm B}} \sum_i |\psi_i(\mathbf{R}_i)|^2\langle\mathbf{I}_i\rangle,
\end{equation}
where $g_{\rm e}^*\approx -0.4$ is the effective g-factor, $\mu_{\rm{B}}$ the Bohr magneton and $A=90\,\upmu$eV the hyperfine coupling strength. For simplicity, we assume a constant hyperfine coupling strength for all species.
The model can be extended straightforwardly to drop this assumption, but this would not change our overall conclusion. Due to the high temperature of the nuclear spin bath, the Overhauser field fluctuates around its average. This field fluctuation can be modelled as a classical, Gaussian distributed noise variable with a standard deviation of $\frac{B_{\rm{max}}}{\sqrt{N}}\approx \SI{4}{\milli\tesla}$ in GaAs quantum dots, where $B_{\rm{max}}$ is the maximum Overhauser field corresponding to full polarization and $N$ the effective number of nuclei in one dot.\cite{Taylor2007}
For $N\gg1$ the Overhauser field can be expressed as 
 \begin{equation}
 \mathbf{B}_{\rm{nuc}}=B_{\rm{max}} \int d^3\mathbf{r}|\psi(\mathbf{r})|^2\mathbf{P}(\mathbf{r},t).
 \end{equation}
Here $\mathbf{P}(\mathbf{r},t)$ is a dimensionless local nuclear polarization density and $B_{\rm{max}}=\frac{ A I} {g_{\rm e}^*\mu_{\rm B}} $ where $I$ denotes the total nuclear spin quantum number. In the following only the component of the Overhauser field aligned with the external magnetic field is evaluated because this is the principal component seen and influenced by the electron. The relative polarization density can then be expressed as $P(\mathbf{r}, t) = \langle I^z_i \rangle / I$.
 
 \subsection{Spin diffusion model}
For $t \gtrsim 100\,\upmu$s nuclear dipole-dipole interactions lead to a diffusion-like redistribution of the local nuclear polarization inside and outside the quantum dot.\cite{Yao2006, Shulman1958} At magnetic fields exceeding $20\,$mT, hyperfine mediated diffusion can be neglected.\cite{Reilly2010} We neglect nuclear spin-lattice relaxation as well because the relaxation time $T_1^{\rm{nuc}}\approx \SI{20}{\minute}$ is  much longer than the typical correlation time of the Overhauser field (seconds to about 1 minute).\cite{Schreiner1997,Reilly2008}
We also neglect the Knight shift, which slightly suppresses diffusion in the region occupied by the electron.

The nuclear dipole-dipole interaction can thus be described by the diffusion equation \cite{Gong2011} on distance scales larger than the lattice spacing and time scales longer than the nearest-neighbour dipole-dipole interaction time. The corresponding 
dynamics of the relative polarization are given by
\begin{equation}\label{eq:spindiff}
\partial_t P(\mathbf{r},t) = D\nabla^2 P(\mathbf{r},t),
\end{equation}
where $D$ is the diffusion constant.\cite{Taylor2006}
The random nuclear fluctuations can be incorporated via Langevin forces. A spatial Fourier transformation of equation \eqref{eq:spindiff} gives
\begin{equation}\label{eq:spindifffft}
\partial_t P^k(t)=-D\mathbf{k}^2P^k(t)+ \Xi^k(t) 
\end{equation}
where $\Xi^k$ is the Langevin force acting on the Fourier mode with wave vector k and satisfies 
\begin{equation}\label{eq:langevin}
\left \langle \Xi^k(t')\Xi^{k'}(t'') \right \rangle =S_{\Xi,k, k'}\delta_{k,k'}\delta(t'-t'').
\end{equation}
Here,  $S_{\Xi,k, k'}$ is the magnitude of the frequency independent noise of the Langevin forces for each Fourier mode.\cite{Taylor2006} 
To obtain the dynamics of the total Overhauser field, we 
rewrite equation \eqref{eq:spindifffft} in terms of the contributions of the Fourier components to the total Overhauser field $B_{\rm{nuc}} = \sum_k B_{\rm{nuc}}^k$ with 
\begin{equation}
B_{\rm{nuc}}^k(t)	= f_k \cdot P^k(t),
\end{equation}
$f_k=B_{\rm{max}} |\psi|_k^2 $ and  $|\psi|_k^2$ denoting the Fourier transform
of  $|\psi|^2$.
This transformation leads to
\begin{equation}\label{eq:diffnodnp}
\frac{dB_{\rm{nuc}}}{dt}= \frac{d}{dt}\sum_k B_{\rm{nuc}}^k(t)=-\sum_k B_{\rm{nuc}}^k(t) D \mathbf{k}^2 - \sum_k f_k \Xi^k(t). 
\end{equation}

If DNP is applied, a term describing the resulting change in polarization has to be added. Pumping introduces a change of $B_{\rm{nuc}}$ by locally polarizing spins. If feedback is used, the pump rate $dB_{\rm{nuc}}/dt$ is a function of instantaneous $B_{\rm{nuc}}$. We linearise this relation by introducing the feedback gain $\gamma_{\rm{pump}}= - \frac{d \left(dB_{\rm{nuc}}/dt\right)}{dB_{\rm{nuc}}}$ and thus obtain
\begin{multline}\label{eq:difftotal}
\frac{d}{dt}\sum_k B_{\rm{nuc}}^k(t)=-\gamma_{\rm{pump}} \sum_k B_{\rm{nuc}}^k(t)\\ -\sum_k B_{\rm{nuc}}^k(t) D k^2 - \sum_k f_k \Xi^k(t).
\end{multline}
As the effective nuclear field is weighted by the Fourier transformed electronic charge density $|\psi|_k^2$, modes with a wave length small in comparison with the spread of the electron wave function are not relevant. The dominant low-$k$ modes on the other hand are slow compared to the feedback time scale since typical decorrelation times of nuclear spin diffusion are seconds
, whereas feedback acts in the sub-second regime.\cite{Reilly2008, Bluhm2010} Thus the term $\sum_k B_{\rm{nuc}}^k D \mathbf{k}^2$ will be small compared to $\gamma_{\rm{pump}}$ and can be neglected for all k-modes. Finally the change in the nuclear magnetic field can be described as
 \begin{equation}
\frac{d}{dt}B_{\rm{nuc}}(t)=-\gamma_{\rm{pump}} B_{\rm{nuc}}(t) - \sum_k f_k \Xi^k(t).
\end{equation}
Solving this first-order inhomogeneous linear differential equation in $B_{\rm{nuc}}(t)$ leads to
\begin{equation*}
B_{\rm{nuc}}(t)=\int\limits_{-\infty}^t dt' e^{-\gamma_{\rm{pump}}(t-t')} \sum_k f_k \Xi^k(t'). 
\end{equation*}
The electronic dephasing time $T_2^*$ depends on the variance of the Overhauser field, 
\begin{eqnarray*}
\langle B_{\rm{nuc}}^2(t)\rangle & = & \int\limits_{-\infty}^t\int\limits_{-\infty}^t dt'dt'' e^{-\gamma_{\rm{pump}}(t-t')}e^{-\gamma_{\rm{pump}}(t-t'')}\\
& &\sum_{k,k'} f_k f_{k'} \langle \Xi^k(t') \Xi^{k'}(t'')\rangle.
\end{eqnarray*}
Substituting $\langle \Xi^k(t')\Xi^{k'}(t'')\rangle$ from equation \eqref{eq:langevin} and carrying out the integrals, we thus find that the variance in the presence of pumping is given by
\begin{equation} \label{eq:nar}
\langle B_{\rm{nuc}}^2(t)\rangle_{\rm{pump}} = \frac{S_{\dot{B}}}{2\gamma_{\rm{pump}}},
\end{equation}
where $S_{\dot{B}}=\sum_k f_k^2 S_{\Xi,k}$ is the spectrum of the rate of change of the Overhauser field. Consequently the Overhauser field variance is directly related to this measure of the strength of the Langevin forces and the feedback gain determined by the pump rate.
Note that as a result of neglecting the $k^2$-terms, this result is of the same
form as previous models not explicitly considering spin diffusion, but with a different expression for $S_{\dot{B}}$.

The feedback gain $\gamma_{\rm{pump}}$ can be measured (e.g. see section \ref{sec:comp}) or computed from a microscopic model. To experimentally obtain $S_{\dot{B}}$, we relate it to the fluctuation spectrum of $B_{\rm{nuc}}$ without DNP, $S_{B_{\rm{nuc}}}(\omega)$,  which can be measured independently. Starting from equation \eqref{eq:diffnodnp} and following Ref. \onlinecite{Taylor2006}, one obtains
\begin{equation}
S_{B_{\rm{nuc}}}(\omega) =\frac{B_{\rm{max}}^2}{N}\int dt \frac{e^{-i\omega t}}{\left(1+t D/\sigma_{xy}^2\right)\sqrt{1+t D/\sigma_z^2}}
\end{equation}

for a Gaussian electron wave function $|\psi(\mathbf{r})|^2=1/\left[(2\pi)^{3/2}\sigma_\perp^2\sigma_z\right]\exp\left[-\left(\mathbf{r}^2_\perp/\sigma_\perp^2+z^2/\sigma_z^2\right)/2\right]$ with $\sigma_z$ as the vertical and $\sigma_\perp$ as the lateral spread of the wave function. 
Approximating the $z$-dependence of the wave function as Gaussion is not expected to introduce major errors because the exact results for all relevant quantum well potentials (triangular, rectangular) also exhibit a quadratic maximum and a fast decay of the wave function, thereby leading to the same qualitative diffusion behaviour.\cite{Witzel2008} 
The resulting noise correlator can be read off as 
\begin{equation}\label{eq:noisecorr}
\langle B_{\rm{nuc}}(t)B_{\rm{nuc}}(0)\rangle_{\rm{eq}}=\frac{B_{\rm{max}}^2}{N}\frac{1}{\left(1+t D/\sigma_{xy}^2\right)\sqrt{1+t D/\sigma_z^2}}.
\end{equation}
On time scales shorter than the correlation time $t_{\rm c}=\min(\frac{\sigma^2_i}{D})\approx\SI{1}{\second}$, the correlator can be approximated to first order as
\begin{eqnarray} 
\langle B_{\rm{nuc}}(t)B_{\rm{nuc}}(0)\rangle_{\rm{eq}}&=&\frac{B_{\rm{max}}^2}{N}\cdot\left[1-\left(D/\sigma_{xy}^2+D/2\sigma_z^2\right)t\right] \nonumber \\
&=& \langle B_{\rm{nuc}}^2(0)\rangle- \frac{1}{2}S_{\dot{B}}t \label{eq:nuccor}.
\end{eqnarray}
The last form is obtained by again neglecting the $k^2$ terms in equation \eqref{eq:diffnodnp}, which is valid for large $\omega$ corresponding to short times $t$.
The reason for this simple result is that different $k$-modes have Lorentzian spectra with different corner frequencies $1/{t_c}$, but on the feedback time scale (shorter than correlation time $t_c$), all relevant modes show an universal $1/\omega^2$ behaviour leading to a linear time dependence of the time domain correlator.

Note that since the different diffusion components have different relaxation times, we cannot relate the narrowing of the Overhauser distribution to the maximal achievable field by DNP.\cite{Reilly2008} In the diffusive case considered here, this maximum field depends on dimension, relaxation rates and diffusion constants. Instead, we employ the directly measurable fluctuation spectrum without DNP. 

While we have only considered a single quantum dot so far for simplicity, our model is also applicable to double quantum dots. For the most widely studied S-T$_0$ qubits, the nuclear field $B_{\rm{nuc}}$ is replaced by the nuclear field gradient between the two dots, $\Delta B_{\rm{nuc}}=B_{\rm{nuc}}^L-B_{\rm{nuc}}^R$.  As one does not expect correlations between the dots, the total spectrum is the sum of those from each dot.

\section{Comparison with experiment} \label{sec:comp}

We now apply our model to the experiment from Ref. \onlinecite{Bluhm2010} performed on a gated double quantum dot in a GaAs/AlGaAs heterostructure. DNP was obtained from Landau-Zener sweeps across the crossing of the singlet state S and the $m = 1$ triplet state T$_+$, where spin angular momentum can be transferred between electrons and nuclei. Feedback was introduced by preceding each sweep with a free evolution of the qubit driven by the current hyperfine field gradient $\Delta B_{\rm{nuc}}^z$. This precession modulates the pump rate as a function of $\Delta B_{\rm{nuc}}^z$ because only one of the two possible qubit states at its end allow spin transfer to the nuclei in the subsequent sweep.

\begin{figure}[htp]
  \centering
      \includegraphics{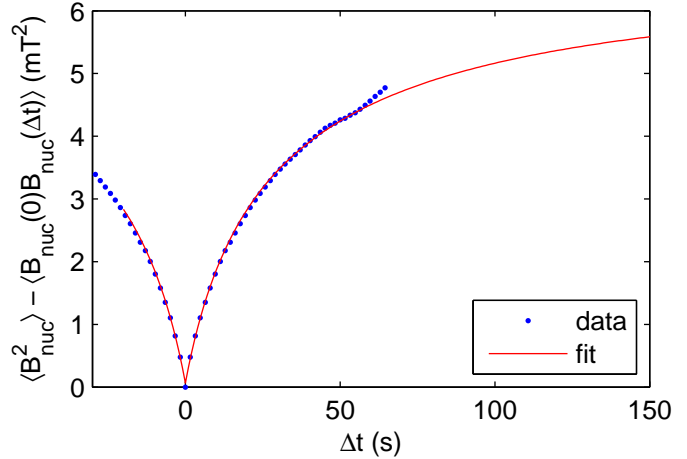}
  \caption[Measured noise correlator of the Overhauser field fluctuations]{Measured noise correlator of the Overhauser field fluctuations in a GaAs double quantum dot without DNP obtained from a time series of  measurements of $\Delta B_{\rm{nuc}}^z$ via FID measurements with a sampling rate of $0.6\,$Hz. 
We plot the difference between the correlator and the variance
because this subtraction cancels a relatively large statistical
uncertainty in the variance arising from the long correlation
time. The short-time behavior of the correlator is not affected by
this uncertainty. The slope near $\rm{\Delta} t=0$ yields the noise magnitude $S_{\dot{B}}=0.53\,\si{\milli\tesla^2/\second}$, which can be used to predict the variance of the Overhauser field distribution. }
  \label{fig:nuccor}
\end{figure}

The independently measured noise correlator of the Overhauser field fluctuations in the same sample is shown in figure \ref{fig:nuccor}. By fitting the noise correlator with equation \eqref{eq:noisecorr}, the noise magnitude $S_{\dot{B}}$ can be calculated according to equation \eqref{eq:nuccor} yielding $S_{\dot{B}}=0.53\,\si{\milli\tesla^2/\second}$. $\sigma_z/\sigma_{xy}$ was fixed at 0.32, and the fitted value 
of $\langle B_{\rm{nuc}}^2\rangle^{1/2} = 2.5\,\si{\milli\tesla}$ agrees with 
typical values for $T_2^*$. With $N = 4.3 \cdot 10^6$ nuclear spins corresponding to  $\sigma_z = 6\,\si{\nano\meter}$, the fit gives a value for the diffusion constant of $D=5\cdot 10^{-14}\,\si{\centi\meter^2/\second}$, which is also reasonable.\cite{Paget1977, Reilly2008} 

The feedback gain $\gamma_{\rm{pump}}$ can be derived from the 
DNP-induced rate of change of the nuclear polarization, which was reported as
$\SI{40}{\milli\tesla/\second}$ in Ref. \onlinecite{Bluhm2010}. As the feedback pulse was only applied for 30 \% of the time, one obtains a mean pump rate of  
$\Gamma_{\rm{pump}}^{\rm{av}}=\SI{12}{\milli\tesla/\second}$.
For the particular feedback implementation considered, the modulation of this
rate with the probability of the free evolution ending in an S state leads to 
$\frac{dB_{\rm{nuc}}}{dt}(\Delta B_{\rm{nuc}}^z, \tau_{\rm{fb}})=\Gamma_{\rm{pump}}^{\rm{av}}\cdot\left(\cos(\omega_{\rm L} \tau_{\rm{fb}})/2+\frac{1}{2}\right)$ with $\omega_{\rm L}=\frac{g_{\rm e}^*\mu_{\rm B} \Delta B_{\rm{nuc}}^z}{\hbar}$ being the electron Larmor frequency and $\tau_{\rm{fb}}$ the duration of the precession between singlet and triplet state which leads to the feedback effect. Hence a feedback gain of $\gamma_{\rm{pump}}=d\dot{B}_{\rm{nuc}}/dB_{\rm{nuc}}=\SI{6.33}{\second^{-1}}$ is obtained with $\tau_{\rm{fb}}=\SI{30}{\nano\second}$. 
Equation \eqref{eq:nar} then leads to an RMS width of the narrowed nuclear field gradient distribution of $\langle B_{\rm{nuc}}^2\rangle_{\rm{pump}}^{1/2} = \SI{0.20}{\milli\tesla}$, corresponding to 
a dephasing time of $T_2^*=\SI{197}{\nano\second}$. This result is
 a factor 2 larger than the measured value of
$T_2^*=94\,\si{\nano\second}$. The discrepancy likely arises partly from 
the fact that polarization and measurement pulses were each applied for 30 to $40\,$ms, which is too long for the approximation of an average pump rate as used here to be very accurate. In addition, the actual pump rate during feedback may have been lower than in the measurement used here.

The noise spectrum is mainly determined by dot geometry and
independent of the feedback scheme. Therefore to increase the
effectiveness of the narrowing of the distribution, the feedback gain and
accordingly the pump rate should be maximized.

\section{Polarization scheme with EDSR on a single quantum dot} \label{sec:polscheme}
We propose a new polarization scheme, the 'EDSR scheme', based on the work of Laird \textit{et al} \cite{Laird2007} to improve the feedback efficiency. Hyperfine-mediated EDSR can be used to control the nuclear magnetic field in the quantum dot as spin angular momentum conservation requires a simultaneous nuclear and electron spin flip. An oscillating electric field shifts the electron wave function in the dot, which in turn creates an effective perpendicular ac magnetic field because the electron experiences an oscillating hyperfine field $B_{\rm{nuc}}^\perp$.\cite{Rashba2008} $B_{\rm{nuc}}^\perp$ drives the electron spin with a Rabi frequency $\Omega_{\rm{R}}$ which depends on the amplitude of the electric field and the hyperfine coupling strength. Competing flipping mechanisms such as spin-orbit coupling decrease the nuclear spin flip probability as no nuclear spin is required to flip. However, by aligning the external magnetic field perpendicular to the direction of electron motion, the spin orbit channel can be suppressed.\cite{Stepanenko2012}

\begin{figure}[htp]
  \centering
   \includegraphics{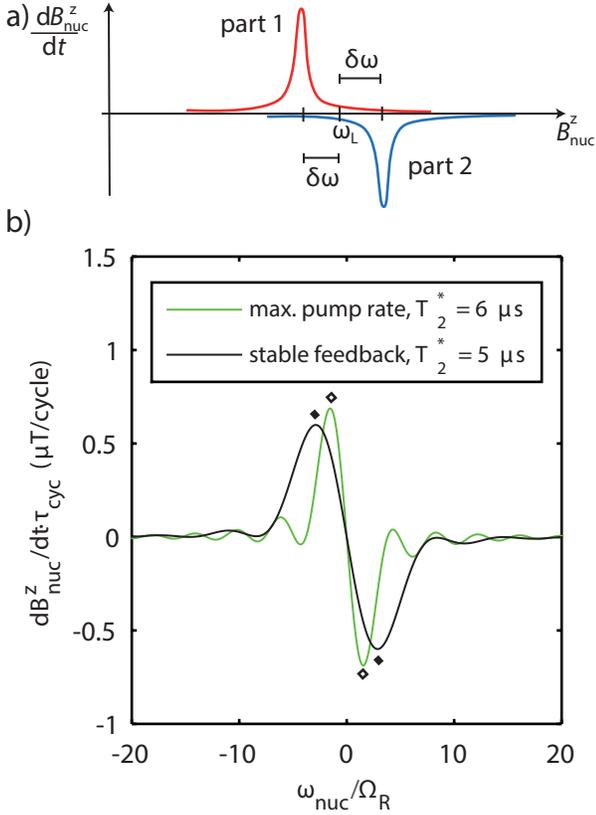}
  \caption[Pump curve of the EDSR feedback scheme]{
  (a) Illustration of the pump curve created by the negatively detuned microwave pulse driving the transition $\ket{\uparrow}\rightarrow\ket{\downarrow}$ (part 1 of the EDSR feedback scheme, red) and the pump curve created by the positively detuned microwave pulse driving $\ket{\downarrow}\rightarrow\ket{\uparrow}$ (part 2 of the EDSR feedback scheme, blue).   (b) Pump curve created by combining the positively and negatively detuned part of the EDSR scheme. In green (black) the maximum-gain (most stable) pump curve yielding a feedback gain $\gamma_{\rm{pump}}=1.2\cdot10^4\ \si{\second^{-1}}$ ($\gamma_{\rm{pump}}=0.8\cdot10^4\ \si{\second^{-1}}$) is depicted. 
  The parameters fulfil the conditions $\frac{\delta\omega}{\Omega_{\rm R}}=1.4$ ($\frac{\delta\omega}{\Omega_{\rm R}}=3$) and $\tau_{\rm{MW}}\delta\omega=2.3$ ($\tau_{\rm{MW}}\delta\omega=\pi$) and the
  exact parameters used are $\Omega_{\rm R}=1.8\,\upmu\si{\second^{-1}}$, $\tau_{\rm{MW}}=0.91\,\si{\micro\second}$ ($\tau_{\rm{MW}}=0.58\,\si{\micro\second}$), $\delta\omega=2.52\,\upmu\si{\second^{-1}}$ ($\delta\omega=5.40\,\upmu\si{\second^{-1}}$) and $\tau_{\rm{cyc}}=\tau_{\rm{MW}}+250\,\si{\nano\second}$. The ratio of the distance between the pair of void (filled) diamonds to the fluctuation width is 12 (19) so that the feedback is stable.  
  }
  \label{fig:pumpmax}
\end{figure}

To stabilize the Overhauser field, two ac electric fields with frequencies each slightly detuned from the desired electron Larmor frequency are sequentially applied to drive transitions between the electron up ($\ket{\uparrow}$) and the electron down ($\ket{\downarrow}$) state in a single quantum dot.
First a $\ket{\uparrow}$ state is initialized and a microwave pulse of duration $\tau_{\rm{MW}}$ at a frequency negatively detuned by $-\delta\omega$ from the targeted electron Larmor frequency $\omega_{\rm L}=g_e^*\mu_{\rm{B}}\frac{B_{\rm{ext}}^z+B_{\rm{nuc}}^z}{\hbar}$ is applied. If the nuclear magnetic field decreases and reaches the value corresponding to this detuned frequency, the transition from $\ket{\uparrow}$ to $\ket{\downarrow}$ is driven. Together with the electron spin a nuclear spin is flipped via the hyperfine interaction and the nuclear magnetic field increases (part 1). The rate of change of the nuclear field as a function of the nuclear field ('pump curve') is illustrated in figure \ref{fig:pumpmax}a. Next the qubit is initialized in a $\ket{\downarrow}$ state and a microwave pulse at a frequency positively detuned by $+\delta\omega$ is applied (part 2). Now the transition from $\ket{\downarrow}$ to $\ket{\uparrow}$ is driven and the nuclear field decreases if the field strength has reached the value corresponding to this positively detuned frequency. The pump curve is also shown in figure \ref{fig:pumpmax}a. 
Alternating between both parts of the scheme leads to a total pump curve with a stable point at the desired nuclear field $\omega_{\rm L}$ as shown in figure \ref{fig:pumpmax}b.  This dependence of the polarization rate on the nuclear state allows the qubit to act as a feedback loop. The negative feedback reduces fluctuations by driving the system to the stable point.

In order to estimate the effectiveness of the EDSR scheme, i.e. the narrowing of the Overhauser distribution, we calculate the pump curve and thus the feedback gain. Firstly the total spin flip probability is evaluated following Ref. \onlinecite{Laird2007}. Because of the parabolic confinement potential, the displacement of the electron wave function is approximately proportional to the electric field. Furthermore,  since the displacement is much smaller than the spread of the electron wave function, the change in nuclear field is approximately proportional to it.  We assume that for each electron spin also a nuclear spin is flipped. Hence the probability distribution of the Rabi frequency arising from the 2D Gaussian distribution of the transverse nuclear field is given by $\rho(\Omega)=(2\Omega/\Omega_{\rm R}^2)\exp(-\Omega^2/\Omega_{\rm R}^2)$, where $\Omega_{\rm R}$ is the typical Rabi frequency. Multiplying the distribution with the EDSR spin-flip probability from an initial $\ket{\downarrow}$ state, we arrive at a total averaged spin flip probability for hyperfine-mediated EDSR of 

\begin{multline} \label{eq:flipprobNorm}
p_{\downarrow}\left(\frac{\Delta\omega}{\Omega_{\rm R}},\tau_{\rm{MW}}\Omega_{\rm R}\right)=\int\limits_0^\infty dx \frac{2x^3}{x^2+\left(\frac{\Delta\omega/2}{\Omega_{\rm R}}\right)^2} \\ \sin^2\left(\sqrt{x^2+\left(\frac{\Delta\omega/2}{\Omega_{\rm R}}\right)^2}\tau_{\rm{MW}}\Omega_{\rm R}\right)e^{-x^2}.
\end{multline}
Here $\Delta\omega$ is the detuning of the microwave from the current Larmor frequency, $\tau_{\rm{MW}}$ the pulse duration and $x=\frac{\Omega}{\Omega_{\rm R}}$ the Rabi frequency relative to the typical Rabi frequency. 
We see that the pump curve depends on the detuning and pulse duration relative to the typical Rabi frequency $\frac{\Delta\omega}{\Omega_{\rm R}}$ and $\tau_{\rm{MW}}\Omega_{\rm R}$, implying that the pump curve and in turn the narrowing is independent of the absolute value of $\Omega_{\rm R}$. The probability $p_\uparrow$ to flip from $\ket{\uparrow}$ to $\ket{\downarrow}$ is given by the same expression. Converting this spin-flip probability to a change in nuclear field and adding up the negatively (part 1) and positively (part 2) detuned pulses results in a pump curve of  
\begin{multline}
\frac{dB_{\rm{nuc}}}{dt}\left(\frac{\omega_{\rm{nuc}}}{\Omega_{\rm R}}, \tau_{\rm{MW}}\Omega_{\rm R}\right)=\frac{1}{\tau_{\rm{cyc}}(\tau_{\rm{MW}})}\frac{B_{\rm{max}}}{N}\cdot\\
 \left[ p_{\downarrow}\left(\frac{\omega_{\rm{nuc}}+\delta\omega}{\Omega_{\rm R}},\tau_{\rm{MW}}\Omega_{\rm R}\right) -p_{\downarrow}\left(\frac{\omega_{\rm{nuc}}-\delta\omega}{\Omega_{\rm R}},\tau_{\rm{MW}}\Omega_{\rm R}\right)\right]
\end{multline}
Here $\tau_{\rm{cyc}}$ is the duration of the feedback cycle and $\omega_{\rm{nuc}}\pm\delta\omega$ the combined detuning of the microwave and the current nuclear field from the desired electron Larmor frequency. Maximizing the feedback gain with respect to $\delta\omega$ and $\tau_{\rm{MW}}$ for Rabi frequencies in the order of MHz (as measured in Ref. \onlinecite{Laird2007}), gives the conditions 
\begin{equation}\label{eq:maxval}
\frac{\delta\omega}{\Omega_{\rm R}}=1.4 \mbox{  and  } \tau_{\rm{MW}}\delta\omega=2.3.
\end{equation}
The resulting pump curve is plotted in figure \ref{fig:pumpmax} in green. For $\tau_{\rm{cyc}}=\tau_{\rm{MW}}$ this pump curve yields the optimal solution for all sets of parameters $\delta\omega$, $\tau_{\rm{MW}}$ and $\Omega_{\rm R}$ fulfilling \eqref{eq:maxval} but if $\tau_{\rm{cyc}}>\tau_{\rm{MW}}$ due to the initialization time, the effective feedback gain still depends slightly on the individual parameters and is not completely fixed by the conditions. 
 Using the largest value from Ref. \onlinecite{Laird2007}, $\Omega_{\rm R}=1.8\,\upmu\si{\second^{-1}}$, for the Rabi frequency leads to the parameters $\tau_{\rm{MW}}=0.91\,\si{\micro\second}$ and $\delta\omega=2.52\,\upmu\si{\second^{-1}}$ and a feedback gain of $\gamma_{\rm{pump}}=3.5\cdot10^4\si{\second}^{-1}$. $\tau_{\rm{cyc}}$ is typically $250\,\si{\nano\second}$ longer than $\tau_{\rm{MW}}$ due to the initialization process. 

In combination with the noise magnitude from figure \ref{fig:nuccor}, the above feedback gain leads to an RMS width of the narrowed nuclear field distribution of $\langle B_{\rm{nuc}}^2\rangle^{1/2}=6.6\cdot 10^{-3}\ \si{\milli\tesla}$ (equation \eqref{eq:nar}) and a dephasing time of $T_2^*=6\,\upmu$s. Thus, the proposed pump scheme promises an improvement of dephasing time by more than two orders of magnitude in comparison to the intrinsic $T_2^*=15\,$ns. The nuclear field gradient variation in turn is larger by a factor of $\sqrt{2}$.

For the feedback to be stable, i.e. to prevent the mean of the distribution from escaping, the nuclear field fluctuations have to be at least one order of magnitude smaller than the the distance of the global maximum and minimum of the pump curve (see distance between $\blacklozenge$ and $\lozenge$ in figure \ref{fig:pumpmax}). Although the narrowing is independent of $\Omega_{\rm R}$, a smaller $\Omega_{\rm R}$ deteriorates the feedback stability. Furthermore, it may be advantageous if the second order local maxima and minima are small, i.e. the pump curve has to be smooth. Appropriate parameters for the latter are $\frac{\delta\omega}{\Omega_{\rm R}}=3$ and $\tau_{\rm{MW}}\delta\omega=\pi$, which give a dephasing time of $T_2^*=5\,\si{\micro\second}$ with $\Omega_{\rm R}=1.8\,\upmu\si{\second^{-1}}$, $\tau_{\rm{MW}}=0.58\,\si{\micro\second}$ and $\delta\omega=5.4\,\upmu\si{\second^{-1}}$. The respective curve is shown in figure \ref{fig:pumpmax} by the black trace. The ratio of peak distance to fluctuation width is $12$ for the maximum (green) curve while it is $19$ for the smoother (black) pump curve. Thus the stabilization criterion is fulfilled for both curves though the smoother curve is more stable.

So far we have only considered polarization noise arising from the diffusive spin dynamics. In addition, shot noise generated by the discrete flipping of spins is a noise source intrinsic to the feedback scheme. The spectral density of this shot noise  to be added to the noise from diffusion $S_{\dot{B}}$ is given by
\begin{equation}
S_{\rm{SN}}=\frac{p_{\rm{flip}}\cdot\left(1-p_{\rm{flip}}\right)}{\tau_{\rm{cyc}}}\cdot\left(\frac{B_{\rm{max}}}{N}\right)^2
\end{equation}
where $\frac{B_{\rm{max}}}{N}$ is the magnetic field change per nuclear spin flip. The remaining factor reflects the variance of the binomially distributed number of flips.  The flip probability at the locking point is $p_{\rm{flip}}=0.41$ for the curve leading to the largest pump rate. With an average cycle time of $\tau_{\rm{cyc}}=1\,\upmu$s, $B_{\rm{max}}=4\,$T and $N=4\cdot10^6$ the noise magnitude is $S_{\rm{SN}}=0.24\,\si{\milli\tesla^2/\second}$. This value is comparable to the diffusion noise. $S_{\rm{SN}}$ has to be added to $S_{\dot{B}}$ in equation \eqref{eq:nar}. Consequently, $T_2^*$ is reduced by a factor of order unity. This shot noise can be reduced by a factor of order unity 
by choosing a smaller $\Omega_{\rm R}$ corresponding to a larger $\tau_{\rm{cyc}}$ and a reduced microwave power without changing the feedback gain $\gamma_{\rm{pump}}$. Ultimately, this reduction is limited by the distance between the peaks of the pump curve becoming comparable to the fluctuation width, which will compromise the stability of the feedback scheme.

\section{Generalization to a double quantum dot and influence of exchange coupling}  \label{sec:dd}

\subsection{Decoupled electrons}

\begin{figure}[htp]
  \centering
      \includegraphics[width=0.5\textwidth]{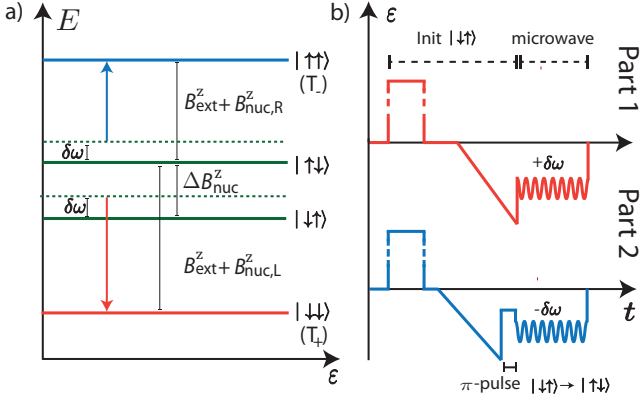}
  \caption[EDSR feedback transitions and feedback pulse sequence]{(a) Energies of the relevant double dot spin states as a function of the detuning in the $(1,1)$ charge state for $\Delta B_{\rm{nuc}}^z> J(\epsilon)$. The detuned microwave signals are indicated by arrows. A drift of $\Delta B_{\rm{nuc}}^z$ brings one closer to and the other further from resonance. (b) Schematic of $\epsilon(t)$ for the positively detuned (top) and negatively detuned (bottom) feedback pulses.}
  \label{fig:pumpsequence}
\end{figure}

The EDSR pump scheme described for a single dot in section \ref{sec:polscheme} can easily be adapted for a double dot system for which the spin states are $\ket{\rm{LR}}$, where L, R$=\{\uparrow,\ \downarrow\}$ refer to the electron state in the left and right dot, respectively. 
In this case, the individual magnetic fields in the dots are stabilized so that the field gradient $\rm{\Delta} B^z_{\rm{nuc}}$ is stabilized as well. 
Initialization of the qubit in $\ket{\downarrow\uparrow}$ ($\ket{\uparrow\downarrow}$) can be achieved by adiabatic preparation from a singlet $S(0,2)$, where both electrons are in the same dot, (and applying a $\pi$-pulse). Consequently both dots are jointly initialized. Then the microwave pulses from section \ref{sec:polscheme} are applied to each individual dot which can be addressed either locally or through frequency selection. For the latter the desired nuclear field gradient $\Delta B_{\rm{nuc}}^z$ has to be large enough so that $\ket{\downarrow\uparrow}$ and $\ket{\uparrow\downarrow}$ are energetically well separated. For the right dot the transitions driven by the microwave are illustrated in figure \ref{fig:pumpsequence}a. The electron is driven into the triplet state $T_+\equiv\ket{\downarrow\downarrow}$ ($T_-\equiv\ket{\uparrow\uparrow}$) with a positively (negatively) detuned microwave pulse if the nuclear field in the right dot has fluctuated downwards (upwards). Consequently the nuclear field in the right dot increases (decreases) dependent on the state of the nuclear field in that dot. The same scheme is applied to the left dot at a different frequency, either simultaneously or sequentially. A sketch of the pulsing scheme is shown in figure \ref{fig:pumpsequence}b. 

\subsection{Exchange coupling}
In a double dot system, an alternative possibility to drive transitions is the exchange coupling $J(\epsilon)$.\cite{Klauser2006} At finite $J$ the microwave pulse leads to an oscillation of the exchange coupling described by $J(t) \approx J(\epsilon_0)+\frac{dJ}{d\epsilon}\delta\epsilon\cos(\omega_{\rm{MW}}t)=J(\epsilon_0)+\delta J\cos(\omega_{\rm{MW}}t)$ where $\delta\epsilon$ is the amplitude of the detuning oscillation dependent on the power of the EDSR burst, which in turn gives the exchange oscillation amplitude $\delta J$. This exchange oscillation can be used as an alternative nuclear spin flipping mechanism. 

Without EDSR the transitions $\ket{S}\rightarrow\ket{T_\pm}$ are driven by the perpendicular magnetic field difference $\rm{\Delta} B_{\rm{nuc}}^\perp$, but as $\rm{\Delta} B_{\rm{nuc}}^\perp\ll$ $ B_z^{\rm{ext}}$ the transition probability is negligible due to conservation of energy. This suppression is overcome by the ac-drive, which corresponds to a Hamiltonian with a time-dependent variation of the level splitting given by
\begin{equation}
H=\frac{\hbar\omega_z\left(t\right)}{2}\sigma_z+\frac{\hbar\omega_\perp}{2}\sigma_y,
\end{equation}
where $\sigma_i$ are the Pauli matrices, $\hbar\omega_z=g_e^*\mu_{\rm B} B_z^{\rm{ext}} - J(t)$ and $\hbar\omega_\perp=g_e^*\mu_{\rm B}B_{\rm{nuc}}^\perp$.
Transforming the eigenbasis for static $\epsilon$ ($\delta\epsilon=\delta J=0$) and neglecting the time dependence of the level splitting in this basis leads to a standard Rabi Hamiltonian with a Rabi frequency of 
\begin{equation}
\Omega_{\rm R}^{J} = \frac{\omega_\perp}{\sqrt{\omega_\perp^2 + \omega_z^2}}\cdot\frac{\delta J}{4\hbar}.
\end{equation}
For a typical external magnetic field $B_z^{\rm{ext}}\gg \rm{\Delta} B_{\rm{nuc}}^\perp$, the Rabi frequency can be approximated to
\begin{equation}
\Omega_{\rm R}^{J} = \frac{\rm{\Delta} B_{\rm{nuc}}^\perp}{B_z^{\rm{ext}}-J\left(\epsilon_0\right)/(g_e^* \mu_{\rm{B}})}\cdot\frac{\delta J}{4\hbar}.
\end{equation}

In order for the exchange flipping mechanism to create a feedback effect, the two eigenstates need to have substantially different components of $\ket{\uparrow\downarrow}$ and $\ket{\downarrow\uparrow}$, thus requiring $J(\epsilon)\lesssim g_e^* \mu_{\rm{B}} \rm{\Delta} B_{\rm{nuc}}^z$. On the other hand, the exchange energy has to be large enough to drive the transitions as the level splitting becomes $J$-independent for $J \ll g_e^* \mu_{\rm{B}}\Delta B_{\rm{nuc}}^z$ and $\delta J$ is bounded by $J(\epsilon)$. 
Inserting typical experimental parameters for $S-T_0$ qubits in GaAs ($\delta J/h=\SI{50}{\mega\hertz}$, $\rm{\Delta} B_{\rm{nuc}}^\perp=\SI{5}{\milli\tesla}$, $B_{z}^{\rm{ext}}=\SI{500}{\milli\tesla})$ yields a Rabi frequency of $\Omega_{\rm R}= 0.81\,\upmu\si{\second^{-1}}$. The Rabi frequency is of the same order of magnitude as the EDSR Rabi frequency. However, a spin-orbit contribution to the $S-T_+$ coupling with a direct $\epsilon$-dependence \cite{Stepanenko2012} may introduce another strong driving channel in this regime, which would have to be avoided carefully.

\section{Summary and conclusion} \label{sec:sum}
In conclusion, we have developed a model that provides a simple relation between the feedback gain of DNP procedures with feedback, the spectrum of nuclear spin fluctuations, and the achievable degree of narrowing of the Overhauser field distribution. The dynamics of the nuclear spins are modelled as a diffusive process. The results are in agreement with earlier experiments. We then proposed a new feedback scheme for gated quantum dots based on EDSR 
which should achieve significantly larger pump rates and thus feedback gains. 
Our model predicts dephasing times of up to $T_2^* = 6\,\upmu$s for this scheme when considering only the diffusive fluctuations of the z-component of the Overhauser field. Interestingly, this value is comparable to the measured value of the Hahn-echo coherence time, which was found to be on the order of $30\,\upmu$s at sufficiently high magnetic fields.\cite{Bluhm2010a} Achieving such long dephasing times would significantly enhance the qubit control fidelity.
In practice, one may however find that effects not considered here impose further limitations. In particular, the faster dynamics of transverse nuclear field components have pronounced effects on Hahn echo amplitudes at applied fields of a few $100\,$mT and below, and are expected to affect FID measurements even more.\cite{Bluhm2010a} Considerably shorter intrinsic FID times have been predicted when taking transverse terms into account.\cite{Yao2006} To partially overcome this limitation, it may turn out important to synchronize both the feedback scheme and measurements to the relative Larmor precession of different spin species. 
On the other hand, it is conceivable that even better results can be obtained at high fields because of an expected cut-off of the spin diffusion spectrum at frequencies exceeding the dipolar nuclear coupling strength. 
In any case, experimentally probing the limits of the proposed EDSR feedback scheme is likely to yield new insights into electron spin dephasing by the nuclear spin bath as it may give access to the so called “intrinsic FID”, which is so far a purely theoretical construct.\cite{Yao2006, Coish2010}

We note $T_2^*$ times for gated dots approaching $3\,\upmu$s have already been demonstrated by a measurement based narrowing procedure.\cite{Shulman2014a} Compared to this approach, our method would have the advantage of achieving a predetermined center of the Overhauser field distribution rather than narrowing around a random value.

\section*{Acknowledgements}
This work was supported by the Alfried Krupp von Bohlen und Halbach Foundation and DFG grant BL 1197/2-1. We thank A. Yacoby, T. Botzem and P. Cerfontaine for discussions.


\begin{thebibliography}{32}%
\makeatletter
\providecommand \@ifxundefined [1]{%
 \@ifx{#1\undefined}
}%
\providecommand \@ifnum [1]{%
 \ifnum #1\expandafter \@firstoftwo
 \else \expandafter \@secondoftwo
 \fi
}%
\providecommand \@ifx [1]{%
 \ifx #1\expandafter \@firstoftwo
 \else \expandafter \@secondoftwo
 \fi
}%
\providecommand \natexlab [1]{#1}%
\providecommand \enquote  [1]{``#1''}%
\providecommand \bibnamefont  [1]{#1}%
\providecommand \bibfnamefont [1]{#1}%
\providecommand \citenamefont [1]{#1}%
\providecommand \href@noop [0]{\@secondoftwo}%
\providecommand \href [0]{\begingroup \@sanitize@url \@href}%
\providecommand \@href[1]{\@@startlink{#1}\@@href}%
\providecommand \@@href[1]{\endgroup#1\@@endlink}%
\providecommand \@sanitize@url [0]{\catcode `\\12\catcode `\$12\catcode
  `\&12\catcode `\#12\catcode `\^12\catcode `\_12\catcode `\%12\relax}%
\providecommand \@@startlink[1]{}%
\providecommand \@@endlink[0]{}%
\providecommand \url  [0]{\begingroup\@sanitize@url \@url }%
\providecommand \@url [1]{\endgroup\@href {#1}{\urlprefix }}%
\providecommand \urlprefix  [0]{URL }%
\providecommand \Eprint [0]{\href }%
\providecommand \doibase [0]{http://dx.doi.org/}%
\providecommand \selectlanguage [0]{\@gobble}%
\providecommand \bibinfo  [0]{\@secondoftwo}%
\providecommand \bibfield  [0]{\@secondoftwo}%
\providecommand \translation [1]{[#1]}%
\providecommand \BibitemOpen [0]{}%
\providecommand \bibitemStop [0]{}%
\providecommand \bibitemNoStop [0]{.\EOS\space}%
\providecommand \EOS [0]{\spacefactor3000\relax}%
\providecommand \BibitemShut  [1]{\csname bibitem#1\endcsname}%
\let\auto@bib@innerbib\@empty
\bibitem [{\citenamefont {Danon}\ \emph {et~al.}(2009)\citenamefont {Danon},
  \citenamefont {Vink}, \citenamefont {Koppens}, \citenamefont {Nowack},
  \citenamefont {Vandersypen},\ and\ \citenamefont {Nazarov}}]{Danon2009a}%
  \BibitemOpen
  \bibfield  {author} {\bibinfo {author} {\bibfnamefont {J.}~\bibnamefont
  {Danon}}, \bibinfo {author} {\bibfnamefont {I.}~\bibnamefont {Vink}},
  \bibinfo {author} {\bibfnamefont {F.}~\bibnamefont {Koppens}}, \bibinfo
  {author} {\bibfnamefont {K.}~\bibnamefont {Nowack}}, \bibinfo {author}
  {\bibfnamefont {L.}~\bibnamefont {Vandersypen}}, \ and\ \bibinfo {author}
  {\bibfnamefont {Y.}~\bibnamefont {Nazarov}},\ }\href {\doibase
  10.1103/PhysRevLett.103.046601} {\bibfield  {journal} {\bibinfo  {journal}
  {Physical Review Letters}\ }\textbf {\bibinfo {volume} {103}},\ \bibinfo
  {pages} {046601} (\bibinfo {year} {2009})}\BibitemShut {NoStop}%
\bibitem [{\citenamefont {Rudner}\ and\ \citenamefont
  {Levitov}(2007)}]{Rudner2007}%
  \BibitemOpen
  \bibfield  {author} {\bibinfo {author} {\bibfnamefont {M.}~\bibnamefont
  {Rudner}}\ and\ \bibinfo {author} {\bibfnamefont {L.}~\bibnamefont
  {Levitov}},\ }\href {\doibase 10.1103/PhysRevLett.99.246602} {\bibfield
  {journal} {\bibinfo  {journal} {Physical Review Letters}\ }\textbf {\bibinfo
  {volume} {99}},\ \bibinfo {pages} {246602} (\bibinfo {year}
  {2007})}\BibitemShut {NoStop}%
\bibitem [{\citenamefont {Xu}\ \emph {et~al.}(2009)\citenamefont {Xu},
  \citenamefont {Yao}, \citenamefont {Sun}, \citenamefont {Steel},
  \citenamefont {Bracker}, \citenamefont {Gammon},\ and\ \citenamefont
  {Sham}}]{Xu2009}%
  \BibitemOpen
  \bibfield  {author} {\bibinfo {author} {\bibfnamefont {X.}~\bibnamefont
  {Xu}}, \bibinfo {author} {\bibfnamefont {W.}~\bibnamefont {Yao}}, \bibinfo
  {author} {\bibfnamefont {B.}~\bibnamefont {Sun}}, \bibinfo {author}
  {\bibfnamefont {D.~G.}\ \bibnamefont {Steel}}, \bibinfo {author}
  {\bibfnamefont {A.~S.}\ \bibnamefont {Bracker}}, \bibinfo {author}
  {\bibfnamefont {D.}~\bibnamefont {Gammon}}, \ and\ \bibinfo {author}
  {\bibfnamefont {L.~J.}\ \bibnamefont {Sham}},\ }\href {\doibase
  10.1038/nature08120} {\bibfield  {journal} {\bibinfo  {journal} {Nature}\
  }\textbf {\bibinfo {volume} {459}},\ \bibinfo {pages} {1105} (\bibinfo {year}
  {2009})}\BibitemShut {NoStop}%
\bibitem [{\citenamefont {Klauser}\ \emph {et~al.}(2006)\citenamefont
  {Klauser}, \citenamefont {Coish},\ and\ \citenamefont {Loss}}]{Klauser2006}%
  \BibitemOpen
  \bibfield  {author} {\bibinfo {author} {\bibfnamefont {D.}~\bibnamefont
  {Klauser}}, \bibinfo {author} {\bibfnamefont {W.}~\bibnamefont {Coish}}, \
  and\ \bibinfo {author} {\bibfnamefont {D.}~\bibnamefont {Loss}},\ }\href
  {\doibase 10.1103/PhysRevB.73.205302} {\bibfield  {journal} {\bibinfo
  {journal} {Physical Review B}\ }\textbf {\bibinfo {volume} {73}},\ \bibinfo
  {pages} {205302} (\bibinfo {year} {2006})}\BibitemShut {NoStop}%
\bibitem [{\citenamefont {Bluhm}\ \emph
  {et~al.}(2010{\natexlab{a}})\citenamefont {Bluhm}, \citenamefont {Foletti},
  \citenamefont {Mahalu}, \citenamefont {Umansky},\ and\ \citenamefont
  {Yacoby}}]{Bluhm2010}%
  \BibitemOpen
  \bibfield  {author} {\bibinfo {author} {\bibfnamefont {H.}~\bibnamefont
  {Bluhm}}, \bibinfo {author} {\bibfnamefont {S.}~\bibnamefont {Foletti}},
  \bibinfo {author} {\bibfnamefont {D.}~\bibnamefont {Mahalu}}, \bibinfo
  {author} {\bibfnamefont {V.}~\bibnamefont {Umansky}}, \ and\ \bibinfo
  {author} {\bibfnamefont {A.}~\bibnamefont {Yacoby}},\ }\href {\doibase
  10.1103/PhysRevLett.105.216803} {\bibfield  {journal} {\bibinfo  {journal}
  {Physical Review Letters}\ }\textbf {\bibinfo {volume} {105}},\ \bibinfo
  {pages} {216803} (\bibinfo {year} {2010}{\natexlab{a}})}\BibitemShut
  {NoStop}%
\bibitem [{\citenamefont {Ladd}\ \emph {et~al.}(2010)\citenamefont {Ladd},
  \citenamefont {Press}, \citenamefont {{De Greve}}, \citenamefont {McMahon},
  \citenamefont {Friess}, \citenamefont {Schneider}, \citenamefont {Kamp},
  \citenamefont {H\"{o}fling}, \citenamefont {Forchel},\ and\ \citenamefont
  {Yamamoto}}]{Ladd2010}%
  \BibitemOpen
  \bibfield  {author} {\bibinfo {author} {\bibfnamefont {T.~D.}\ \bibnamefont
  {Ladd}}, \bibinfo {author} {\bibfnamefont {D.}~\bibnamefont {Press}},
  \bibinfo {author} {\bibfnamefont {K.}~\bibnamefont {{De Greve}}}, \bibinfo
  {author} {\bibfnamefont {P.~L.}\ \bibnamefont {McMahon}}, \bibinfo {author}
  {\bibfnamefont {B.}~\bibnamefont {Friess}}, \bibinfo {author} {\bibfnamefont
  {C.}~\bibnamefont {Schneider}}, \bibinfo {author} {\bibfnamefont
  {M.}~\bibnamefont {Kamp}}, \bibinfo {author} {\bibfnamefont {S.}~\bibnamefont
  {H\"{o}fling}}, \bibinfo {author} {\bibfnamefont {A.}~\bibnamefont
  {Forchel}}, \ and\ \bibinfo {author} {\bibfnamefont {Y.}~\bibnamefont
  {Yamamoto}},\ }\href {\doibase 10.1103/PhysRevLett.105.107401} {\bibfield
  {journal} {\bibinfo  {journal} {Physical Review Letters}\ }\textbf {\bibinfo
  {volume} {105}},\ \bibinfo {pages} {107401} (\bibinfo {year}
  {2010})}\BibitemShut {NoStop}%
\bibitem [{\citenamefont {H\"{o}gele}\ \emph {et~al.}(2012)\citenamefont
  {H\"{o}gele}, \citenamefont {Kroner}, \citenamefont {Latta}, \citenamefont
  {Claassen}, \citenamefont {Carusotto}, \citenamefont {Bulutay},\ and\
  \citenamefont {Imamoglu}}]{Hogele2012}%
  \BibitemOpen
  \bibfield  {author} {\bibinfo {author} {\bibfnamefont {A.}~\bibnamefont
  {H\"{o}gele}}, \bibinfo {author} {\bibfnamefont {M.}~\bibnamefont {Kroner}},
  \bibinfo {author} {\bibfnamefont {C.}~\bibnamefont {Latta}}, \bibinfo
  {author} {\bibfnamefont {M.}~\bibnamefont {Claassen}}, \bibinfo {author}
  {\bibfnamefont {I.}~\bibnamefont {Carusotto}}, \bibinfo {author}
  {\bibfnamefont {C.}~\bibnamefont {Bulutay}}, \ and\ \bibinfo {author}
  {\bibfnamefont {A.}~\bibnamefont {Imamoglu}},\ }\href {\doibase
  10.1103/PhysRevLett.108.197403} {\bibfield  {journal} {\bibinfo  {journal}
  {Physical Review Letters}\ }\textbf {\bibinfo {volume} {108}},\ \bibinfo
  {pages} {197403} (\bibinfo {year} {2012})}\BibitemShut {NoStop}%
\bibitem [{\citenamefont {Shulman}\ \emph {et~al.}(2014)\citenamefont
  {Shulman}, \citenamefont {Harvey}, \citenamefont {Nichol}, \citenamefont
  {Bartlett}, \citenamefont {Doherty}, \citenamefont {Umansky},\ and\
  \citenamefont {Yacoby}}]{Shulman2014a}%
  \BibitemOpen
  \bibfield  {author} {\bibinfo {author} {\bibfnamefont {M.~D.}\ \bibnamefont
  {Shulman}}, \bibinfo {author} {\bibfnamefont {S.~P.}\ \bibnamefont {Harvey}},
  \bibinfo {author} {\bibfnamefont {J.~M.}\ \bibnamefont {Nichol}}, \bibinfo
  {author} {\bibfnamefont {S.~D.}\ \bibnamefont {Bartlett}}, \bibinfo {author}
  {\bibfnamefont {A.~C.}\ \bibnamefont {Doherty}}, \bibinfo {author}
  {\bibfnamefont {V.}~\bibnamefont {Umansky}}, \ and\ \bibinfo {author}
  {\bibfnamefont {A.}~\bibnamefont {Yacoby}},\ }\href {\doibase
  10.1038/ncomms6156} {\bibfield  {journal} {\bibinfo  {journal} {Nature
  communications}\ }\textbf {\bibinfo {volume} {5}},\ \bibinfo {pages} {5156}
  (\bibinfo {year} {2014})}\BibitemShut {NoStop}%
\bibitem [{\citenamefont {Latta}\ \emph {et~al.}(2009)\citenamefont {Latta},
  \citenamefont {H\"{o}gele}, \citenamefont {Zhao}, \citenamefont {Vamivakas},
  \citenamefont {Maletinsky}, \citenamefont {Kroner}, \citenamefont {Dreiser},
  \citenamefont {Carusotto}, \citenamefont {Badolato}, \citenamefont {Schuh},
  \citenamefont {Wegscheider}, \citenamefont {Atature},\ and\ \citenamefont
  {Imamoglu}}]{Latta2009}%
  \BibitemOpen
  \bibfield  {author} {\bibinfo {author} {\bibfnamefont {C.}~\bibnamefont
  {Latta}}, \bibinfo {author} {\bibfnamefont {A.}~\bibnamefont {H\"{o}gele}},
  \bibinfo {author} {\bibfnamefont {Y.}~\bibnamefont {Zhao}}, \bibinfo {author}
  {\bibfnamefont {A.~N.}\ \bibnamefont {Vamivakas}}, \bibinfo {author}
  {\bibfnamefont {P.}~\bibnamefont {Maletinsky}}, \bibinfo {author}
  {\bibfnamefont {M.}~\bibnamefont {Kroner}}, \bibinfo {author} {\bibfnamefont
  {J.}~\bibnamefont {Dreiser}}, \bibinfo {author} {\bibfnamefont
  {I.}~\bibnamefont {Carusotto}}, \bibinfo {author} {\bibfnamefont
  {A.}~\bibnamefont {Badolato}}, \bibinfo {author} {\bibfnamefont
  {D.}~\bibnamefont {Schuh}}, \bibinfo {author} {\bibfnamefont
  {W.}~\bibnamefont {Wegscheider}}, \bibinfo {author} {\bibfnamefont
  {M.}~\bibnamefont {Atature}}, \ and\ \bibinfo {author} {\bibfnamefont
  {A.}~\bibnamefont {Imamoglu}},\ }\href {\doibase 10.1038/nphys1363}
  {\bibfield  {journal} {\bibinfo  {journal} {Nature Physics}\ }\textbf
  {\bibinfo {volume} {5}},\ \bibinfo {pages} {758} (\bibinfo {year}
  {2009})}\BibitemShut {NoStop}%
\bibitem [{\citenamefont {Greilich}\ \emph {et~al.}(2007)\citenamefont
  {Greilich}, \citenamefont {Shabaev}, \citenamefont {Yakovlev}, \citenamefont
  {Efros}, \citenamefont {Yugova}, \citenamefont {Reuter}, \citenamefont
  {Wieck},\ and\ \citenamefont {Bayer}}]{Greilich2007}%
  \BibitemOpen
  \bibfield  {author} {\bibinfo {author} {\bibfnamefont {A.}~\bibnamefont
  {Greilich}}, \bibinfo {author} {\bibfnamefont {A.}~\bibnamefont {Shabaev}},
  \bibinfo {author} {\bibfnamefont {D.~R.}\ \bibnamefont {Yakovlev}}, \bibinfo
  {author} {\bibfnamefont {A.~L.}\ \bibnamefont {Efros}}, \bibinfo {author}
  {\bibfnamefont {I.~A.}\ \bibnamefont {Yugova}}, \bibinfo {author}
  {\bibfnamefont {D.}~\bibnamefont {Reuter}}, \bibinfo {author} {\bibfnamefont
  {A.~D.}\ \bibnamefont {Wieck}}, \ and\ \bibinfo {author} {\bibfnamefont
  {M.}~\bibnamefont {Bayer}},\ }\href {\doibase 10.1126/science.1146850}
  {\bibfield  {journal} {\bibinfo  {journal} {Science}\ }\textbf {\bibinfo
  {volume} {317}},\ \bibinfo {pages} {1896} (\bibinfo {year}
  {2007})}\BibitemShut {NoStop}%
\bibitem [{\citenamefont {Vink}\ \emph {et~al.}(2009)\citenamefont {Vink},
  \citenamefont {Nowack}, \citenamefont {Koppens}, \citenamefont {Danon},
  \citenamefont {Nazarov},\ and\ \citenamefont {Vandersypen}}]{Vink2009}%
  \BibitemOpen
  \bibfield  {author} {\bibinfo {author} {\bibfnamefont {I.~T.}\ \bibnamefont
  {Vink}}, \bibinfo {author} {\bibfnamefont {K.~C.}\ \bibnamefont {Nowack}},
  \bibinfo {author} {\bibfnamefont {F.~H.~L.}\ \bibnamefont {Koppens}},
  \bibinfo {author} {\bibfnamefont {J.}~\bibnamefont {Danon}}, \bibinfo
  {author} {\bibfnamefont {Y.~V.}\ \bibnamefont {Nazarov}}, \ and\ \bibinfo
  {author} {\bibfnamefont {L.~M.~K.}\ \bibnamefont {Vandersypen}},\ }\href
  {\doibase 10.1038/nphys1366} {\bibfield  {journal} {\bibinfo  {journal}
  {Nature Physics}\ }\textbf {\bibinfo {volume} {5}},\ \bibinfo {pages} {764}
  (\bibinfo {year} {2009})}\BibitemShut {NoStop}%
\bibitem [{\citenamefont {Yang}\ and\ \citenamefont {Sham}(2013)}]{Yang2013}%
  \BibitemOpen
  \bibfield  {author} {\bibinfo {author} {\bibfnamefont {W.}~\bibnamefont
  {Yang}}\ and\ \bibinfo {author} {\bibfnamefont {L.~J.}\ \bibnamefont
  {Sham}},\ }\href {\doibase 10.1103/PhysRevB.88.235304} {\bibfield  {journal}
  {\bibinfo  {journal} {Physical Review B}\ }\textbf {\bibinfo {volume} {88}},\
  \bibinfo {pages} {235304} (\bibinfo {year} {2013})}\BibitemShut {NoStop}%
\bibitem [{\citenamefont {Barnes}\ \emph {et~al.}(2012)\citenamefont {Barnes},
  \citenamefont {Cywinski},\ and\ \citenamefont {{Das Sarma}}}]{Barnes2012}%
  \BibitemOpen
  \bibfield  {author} {\bibinfo {author} {\bibfnamefont {E.}~\bibnamefont
  {Barnes}}, \bibinfo {author} {\bibfnamefont {L.}~\bibnamefont {Cywinski}}, \
  and\ \bibinfo {author} {\bibfnamefont {S.}~\bibnamefont {{Das Sarma}}},\
  }\href {\doibase 10.1103/PhysRevLett.109.140403} {\bibfield  {journal}
  {\bibinfo  {journal} {Physical Review Letters}\ }\textbf {\bibinfo {volume}
  {109}},\ \bibinfo {pages} {140403} (\bibinfo {year} {2012})}\BibitemShut
  {NoStop}%
\bibitem [{\citenamefont {Economou}\ and\ \citenamefont
  {Barnes}(2014)}]{Economou2014}%
  \BibitemOpen
  \bibfield  {author} {\bibinfo {author} {\bibfnamefont {S.~E.}\ \bibnamefont
  {Economou}}\ and\ \bibinfo {author} {\bibfnamefont {E.}~\bibnamefont
  {Barnes}},\ }\href {\doibase 10.1103/PhysRevB.89.165301} {\bibfield
  {journal} {\bibinfo  {journal} {Physical Review B}\ }\textbf {\bibinfo
  {volume} {89}},\ \bibinfo {pages} {165301} (\bibinfo {year}
  {2014})}\BibitemShut {NoStop}%
\bibitem [{\citenamefont {Urbaszek}\ \emph {et~al.}(2013)\citenamefont
  {Urbaszek}, \citenamefont {Marie}, \citenamefont {Amand}, \citenamefont
  {Krebs}, \citenamefont {Voisin}, \citenamefont {Maletinsky}, \citenamefont
  {H\"{o}gele},\ and\ \citenamefont {Imamoglu}}]{Urbaszek2013}%
  \BibitemOpen
  \bibfield  {author} {\bibinfo {author} {\bibfnamefont {B.}~\bibnamefont
  {Urbaszek}}, \bibinfo {author} {\bibfnamefont {X.}~\bibnamefont {Marie}},
  \bibinfo {author} {\bibfnamefont {T.}~\bibnamefont {Amand}}, \bibinfo
  {author} {\bibfnamefont {O.}~\bibnamefont {Krebs}}, \bibinfo {author}
  {\bibfnamefont {P.}~\bibnamefont {Voisin}}, \bibinfo {author} {\bibfnamefont
  {P.}~\bibnamefont {Maletinsky}}, \bibinfo {author} {\bibfnamefont
  {A.}~\bibnamefont {H\"{o}gele}}, \ and\ \bibinfo {author} {\bibfnamefont
  {A.}~\bibnamefont {Imamoglu}},\ }\href {\doibase 10.1103/RevModPhys.85.79}
  {\bibfield  {journal} {\bibinfo  {journal} {Rev. Mod. Phys.}\ }\textbf
  {\bibinfo {volume} {85}},\ \bibinfo {pages} {79} (\bibinfo {year}
  {2013})}\BibitemShut {NoStop}%
\bibitem [{\citenamefont {Yang}\ and\ \citenamefont {Sham}(2012)}]{Yang2012}%
  \BibitemOpen
  \bibfield  {author} {\bibinfo {author} {\bibfnamefont {W.}~\bibnamefont
  {Yang}}\ and\ \bibinfo {author} {\bibfnamefont {L.~J.}\ \bibnamefont
  {Sham}},\ }\href {\doibase 10.1103/PhysRevB.85.235319} {\bibfield  {journal}
  {\bibinfo  {journal} {Physical Review B}\ }\textbf {\bibinfo {volume} {85}},\
  \bibinfo {pages} {235319} (\bibinfo {year} {2012})}\BibitemShut {NoStop}%
\bibitem [{\citenamefont {Reilly}\ \emph {et~al.}(2008)\citenamefont {Reilly},
  \citenamefont {Taylor}, \citenamefont {Laird}, \citenamefont {Petta},
  \citenamefont {Marcus}, \citenamefont {Hanson},\ and\ \citenamefont
  {Gossard}}]{Reilly2008}%
  \BibitemOpen
  \bibfield  {author} {\bibinfo {author} {\bibfnamefont {D.~J.}\ \bibnamefont
  {Reilly}}, \bibinfo {author} {\bibfnamefont {J.~M.}\ \bibnamefont {Taylor}},
  \bibinfo {author} {\bibfnamefont {E.~A.}\ \bibnamefont {Laird}}, \bibinfo
  {author} {\bibfnamefont {J.~R.}\ \bibnamefont {Petta}}, \bibinfo {author}
  {\bibfnamefont {C.}~\bibnamefont {Marcus}}, \bibinfo {author} {\bibfnamefont
  {M.~P.}\ \bibnamefont {Hanson}}, \ and\ \bibinfo {author} {\bibfnamefont
  {A.~C.}\ \bibnamefont {Gossard}},\ }\href {\doibase
  10.1103/PhysRevLett.101.236803} {\bibfield  {journal} {\bibinfo  {journal}
  {Physical Review Letters}\ }\textbf {\bibinfo {volume} {101}},\ \bibinfo
  {pages} {236803} (\bibinfo {year} {2008})}\BibitemShut {NoStop}%
\bibitem [{\citenamefont {Sallen}\ \emph {et~al.}(2014)\citenamefont {Sallen},
  \citenamefont {Kunz}, \citenamefont {Amand}, \citenamefont {Bouet},
  \citenamefont {Kuroda}, \citenamefont {Mano}, \citenamefont {Paget},
  \citenamefont {Krebs}, \citenamefont {Marie}, \citenamefont {Sakoda},\ and\
  \citenamefont {Urbaszek}}]{Sallen2014}%
  \BibitemOpen
  \bibfield  {author} {\bibinfo {author} {\bibfnamefont {G.}~\bibnamefont
  {Sallen}}, \bibinfo {author} {\bibfnamefont {S.}~\bibnamefont {Kunz}},
  \bibinfo {author} {\bibfnamefont {T.}~\bibnamefont {Amand}}, \bibinfo
  {author} {\bibfnamefont {L.}~\bibnamefont {Bouet}}, \bibinfo {author}
  {\bibfnamefont {T.}~\bibnamefont {Kuroda}}, \bibinfo {author} {\bibfnamefont
  {T.}~\bibnamefont {Mano}}, \bibinfo {author} {\bibfnamefont {D.}~\bibnamefont
  {Paget}}, \bibinfo {author} {\bibfnamefont {O.}~\bibnamefont {Krebs}},
  \bibinfo {author} {\bibfnamefont {X.}~\bibnamefont {Marie}}, \bibinfo
  {author} {\bibfnamefont {K.}~\bibnamefont {Sakoda}}, \ and\ \bibinfo {author}
  {\bibfnamefont {B.}~\bibnamefont {Urbaszek}},\ }\href {\doibase
  10.1038/ncomms4268} {\bibfield  {journal} {\bibinfo  {journal} {Nature
  communications}\ }\textbf {\bibinfo {volume} {5}},\ \bibinfo {pages} {3268}
  (\bibinfo {year} {2014})}\BibitemShut {NoStop}%
\bibitem [{\citenamefont {Laird}\ \emph {et~al.}(2007)\citenamefont {Laird},
  \citenamefont {Barthel}, \citenamefont {Rashba}, \citenamefont {Marcus},
  \citenamefont {Hanson},\ and\ \citenamefont {Gossard}}]{Laird2007}%
  \BibitemOpen
  \bibfield  {author} {\bibinfo {author} {\bibfnamefont {E.}~\bibnamefont
  {Laird}}, \bibinfo {author} {\bibfnamefont {C.}~\bibnamefont {Barthel}},
  \bibinfo {author} {\bibfnamefont {E.}~\bibnamefont {Rashba}}, \bibinfo
  {author} {\bibfnamefont {C.}~\bibnamefont {Marcus}}, \bibinfo {author}
  {\bibfnamefont {M.}~\bibnamefont {Hanson}}, \ and\ \bibinfo {author}
  {\bibfnamefont {A.}~\bibnamefont {Gossard}},\ }\href {\doibase
  10.1103/PhysRevLett.99.246601} {\bibfield  {journal} {\bibinfo  {journal}
  {Physical Review Letters}\ }\textbf {\bibinfo {volume} {99}},\ \bibinfo
  {pages} {246601} (\bibinfo {year} {2007})}\BibitemShut {NoStop}%
\bibitem [{\citenamefont {Taylor}\ \emph {et~al.}(2007)\citenamefont {Taylor},
  \citenamefont {Petta}, \citenamefont {Johnson}, \citenamefont {Yacoby},
  \citenamefont {Marcus},\ and\ \citenamefont {Lukin}}]{Taylor2007}%
  \BibitemOpen
  \bibfield  {author} {\bibinfo {author} {\bibfnamefont {J.}~\bibnamefont
  {Taylor}}, \bibinfo {author} {\bibfnamefont {J.}~\bibnamefont {Petta}},
  \bibinfo {author} {\bibfnamefont {A.}~\bibnamefont {Johnson}}, \bibinfo
  {author} {\bibfnamefont {A.}~\bibnamefont {Yacoby}}, \bibinfo {author}
  {\bibfnamefont {C.}~\bibnamefont {Marcus}}, \ and\ \bibinfo {author}
  {\bibfnamefont {M.}~\bibnamefont {Lukin}},\ }\href {\doibase
  10.1103/PhysRevB.76.035315} {\bibfield  {journal} {\bibinfo  {journal}
  {Physical Review B}\ }\textbf {\bibinfo {volume} {76}},\ \bibinfo {pages}
  {35315} (\bibinfo {year} {2007})}\BibitemShut {NoStop}%
\bibitem [{\citenamefont {Yao}\ \emph {et~al.}(2006)\citenamefont {Yao},
  \citenamefont {Liu},\ and\ \citenamefont {Sham}}]{Yao2006}%
  \BibitemOpen
  \bibfield  {author} {\bibinfo {author} {\bibfnamefont {W.}~\bibnamefont
  {Yao}}, \bibinfo {author} {\bibfnamefont {R.-B.}\ \bibnamefont {Liu}}, \ and\
  \bibinfo {author} {\bibfnamefont {L.}~\bibnamefont {Sham}},\ }\href {\doibase
  10.1103/PhysRevB.74.195301} {\bibfield  {journal} {\bibinfo  {journal}
  {Physical Review B}\ }\textbf {\bibinfo {volume} {74}},\ \bibinfo {pages}
  {195301} (\bibinfo {year} {2006})}\BibitemShut {NoStop}%
\bibitem [{\citenamefont {Shulman}\ \emph {et~al.}(1958)\citenamefont
  {Shulman}, \citenamefont {Wylunda},\ and\ \citenamefont
  {Hrostowski}}]{Shulman1958}%
  \BibitemOpen
  \bibfield  {author} {\bibinfo {author} {\bibfnamefont {R.}~\bibnamefont
  {Shulman}}, \bibinfo {author} {\bibfnamefont {B.}~\bibnamefont {Wylunda}}, \
  and\ \bibinfo {author} {\bibfnamefont {H.}~\bibnamefont {Hrostowski}},\
  }\href@noop {} {\bibfield  {journal} {\bibinfo  {journal} {Physical Review}\
  }\textbf {\bibinfo {volume} {109}},\ \bibinfo {pages} {808} (\bibinfo {year}
  {1958})}\BibitemShut {NoStop}%
\bibitem [{\citenamefont {Reilly}\ \emph {et~al.}(2010)\citenamefont {Reilly},
  \citenamefont {Taylor}, \citenamefont {Petta}, \citenamefont {Marcus},
  \citenamefont {Hanson},\ and\ \citenamefont {Gossard}}]{Reilly2010}%
  \BibitemOpen
  \bibfield  {author} {\bibinfo {author} {\bibfnamefont {D.~J.}\ \bibnamefont
  {Reilly}}, \bibinfo {author} {\bibfnamefont {J.~M.}\ \bibnamefont {Taylor}},
  \bibinfo {author} {\bibfnamefont {J.~R.}\ \bibnamefont {Petta}}, \bibinfo
  {author} {\bibfnamefont {C.~M.}\ \bibnamefont {Marcus}}, \bibinfo {author}
  {\bibfnamefont {M.~P.}\ \bibnamefont {Hanson}}, \ and\ \bibinfo {author}
  {\bibfnamefont {A.~C.}\ \bibnamefont {Gossard}},\ }\href {\doibase
  10.1103/PhysRevLett.104.236802} {\bibfield  {journal} {\bibinfo  {journal}
  {Physical Review Letters}\ }\textbf {\bibinfo {volume} {104}},\ \bibinfo
  {pages} {236802} (\bibinfo {year} {2010})}\BibitemShut {NoStop}%
\bibitem [{\citenamefont {Schreiner}\ \emph {et~al.}(1997)\citenamefont
  {Schreiner}, \citenamefont {Pascher}, \citenamefont {Denninger},
  \citenamefont {Studenikin}, \citenamefont {Weimann},\ and\ \citenamefont
  {L\"osch}}]{Schreiner1997}%
  \BibitemOpen
  \bibfield  {author} {\bibinfo {author} {\bibfnamefont {M.}~\bibnamefont
  {Schreiner}}, \bibinfo {author} {\bibfnamefont {H.}~\bibnamefont {Pascher}},
  \bibinfo {author} {\bibfnamefont {G.}~\bibnamefont {Denninger}}, \bibinfo
  {author} {\bibfnamefont {S.}~\bibnamefont {Studenikin}}, \bibinfo {author}
  {\bibfnamefont {G.}~\bibnamefont {Weimann}}, \ and\ \bibinfo {author}
  {\bibfnamefont {R.}~\bibnamefont {L\"osch}},\ }\href@noop {} {\bibfield
  {journal} {\bibinfo  {journal} {Solid State Communications}\ }\textbf
  {\bibinfo {volume} {102}},\ \bibinfo {pages} {715} (\bibinfo {year}
  {1997})}\BibitemShut {NoStop}%
\bibitem [{\citenamefont {Gong}\ \emph {et~al.}(2011)\citenamefont {Gong},
  \citenamefont {qi~Yin},\ and\ \citenamefont {Duan}}]{Gong2011}%
  \BibitemOpen
  \bibfield  {author} {\bibinfo {author} {\bibfnamefont {Z.-X.}\ \bibnamefont
  {Gong}}, \bibinfo {author} {\bibfnamefont {Z.}~\bibnamefont {qi~Yin}}, \ and\
  \bibinfo {author} {\bibfnamefont {L.-M.}\ \bibnamefont {Duan}},\ }\href@noop
  {} {\bibfield  {journal} {\bibinfo  {journal} {New Journal of Physics}\
  }\textbf {\bibinfo {volume} {13}},\ \bibinfo {pages} {033036} (\bibinfo
  {year} {2011})}\BibitemShut {NoStop}%
\bibitem [{\citenamefont {Taylor}(2006)}]{Taylor2006}%
  \BibitemOpen
  \bibfield  {author} {\bibinfo {author} {\bibfnamefont {J.~M.}\ \bibnamefont
  {Taylor}},\ }\emph {\bibinfo {title} {{Hyperfine interactions and quantum
  information processing in quantum dots}}},\ \href@noop {} {Ph.D. thesis},\
  \bibinfo  {school} {Harvard University} (\bibinfo {year} {2006})\BibitemShut
  {NoStop}%
\bibitem [{\citenamefont {Witzel}\ and\ \citenamefont {{Das
  Sarma}}(2008)}]{Witzel2008}%
  \BibitemOpen
  \bibfield  {author} {\bibinfo {author} {\bibfnamefont {W.~M.}\ \bibnamefont
  {Witzel}}\ and\ \bibinfo {author} {\bibfnamefont {S.}~\bibnamefont {{Das
  Sarma}}},\ }\href {\doibase 10.1103/PhysRevB.77.165319} {\bibfield  {journal}
  {\bibinfo  {journal} {Physical Review B}\ }\textbf {\bibinfo {volume} {77}},\
  \bibinfo {pages} {165319} (\bibinfo {year} {2008})}\BibitemShut {NoStop}%
\bibitem [{\citenamefont {Paget}(1977)}]{Paget1977}%
  \BibitemOpen
  \bibfield  {author} {\bibinfo {author} {\bibfnamefont {D.}~\bibnamefont
  {Paget}},\ }\href@noop {} {\bibfield  {journal} {\bibinfo  {journal}
  {Physical Review B}\ }\textbf {\bibinfo {volume} {15}},\ \bibinfo {pages}
  {5780} (\bibinfo {year} {1977})}\BibitemShut {NoStop}%
\bibitem [{\citenamefont {Rashba}(2008)}]{Rashba2008}%
  \BibitemOpen
  \bibfield  {author} {\bibinfo {author} {\bibfnamefont {E.}~\bibnamefont
  {Rashba}},\ }\href {\doibase 10.1103/PhysRevB.78.195302} {\bibfield
  {journal} {\bibinfo  {journal} {Physical Review B}\ }\textbf {\bibinfo
  {volume} {78}},\ \bibinfo {pages} {195302} (\bibinfo {year}
  {2008})}\BibitemShut {NoStop}%
\bibitem [{\citenamefont {Stepanenko}\ \emph {et~al.}(2012)\citenamefont
  {Stepanenko}, \citenamefont {Rudner}, \citenamefont {Halperin},\ and\
  \citenamefont {Loss}}]{Stepanenko2012}%
  \BibitemOpen
  \bibfield  {author} {\bibinfo {author} {\bibfnamefont {D.}~\bibnamefont
  {Stepanenko}}, \bibinfo {author} {\bibfnamefont {M.}~\bibnamefont {Rudner}},
  \bibinfo {author} {\bibfnamefont {B.~I.}\ \bibnamefont {Halperin}}, \ and\
  \bibinfo {author} {\bibfnamefont {D.}~\bibnamefont {Loss}},\ }\href {\doibase
  10.1103/PhysRevB.85.075416} {\bibfield  {journal} {\bibinfo  {journal}
  {Physical Review B}\ }\textbf {\bibinfo {volume} {85}},\ \bibinfo {pages}
  {75416} (\bibinfo {year} {2012})}\BibitemShut {NoStop}%
\bibitem [{\citenamefont {Bluhm}\ \emph
  {et~al.}(2010{\natexlab{b}})\citenamefont {Bluhm}, \citenamefont {Foletti},
  \citenamefont {Neder}, \citenamefont {Rudner}, \citenamefont {Mahalu},
  \citenamefont {Umansky},\ and\ \citenamefont {Yacoby}}]{Bluhm2010a}%
  \BibitemOpen
  \bibfield  {author} {\bibinfo {author} {\bibfnamefont {H.}~\bibnamefont
  {Bluhm}}, \bibinfo {author} {\bibfnamefont {S.}~\bibnamefont {Foletti}},
  \bibinfo {author} {\bibfnamefont {I.}~\bibnamefont {Neder}}, \bibinfo
  {author} {\bibfnamefont {M.}~\bibnamefont {Rudner}}, \bibinfo {author}
  {\bibfnamefont {D.}~\bibnamefont {Mahalu}}, \bibinfo {author} {\bibfnamefont
  {V.}~\bibnamefont {Umansky}}, \ and\ \bibinfo {author} {\bibfnamefont
  {A.}~\bibnamefont {Yacoby}},\ }\href {\doibase 10.1038/nphys1856} {\bibfield
  {journal} {\bibinfo  {journal} {Nature Physics}\ }\textbf {\bibinfo {volume}
  {7}},\ \bibinfo {pages} {109} (\bibinfo {year}
  {2010}{\natexlab{b}})}\BibitemShut {NoStop}%
\bibitem [{\citenamefont {Coish}\ \emph {et~al.}(2010)\citenamefont {Coish},
  \citenamefont {Fischer},\ and\ \citenamefont {Loss}}]{Coish2010}%
  \BibitemOpen
  \bibfield  {author} {\bibinfo {author} {\bibfnamefont {W.~A.}\ \bibnamefont
  {Coish}}, \bibinfo {author} {\bibfnamefont {J.}~\bibnamefont {Fischer}}, \
  and\ \bibinfo {author} {\bibfnamefont {D.}~\bibnamefont {Loss}},\ }\href
  {\doibase 10.1103/PhysRevB.81.165315} {\bibfield  {journal} {\bibinfo
  {journal} {Physical Review B}\ }\textbf {\bibinfo {volume} {81}},\ \bibinfo
  {pages} {165315} (\bibinfo {year} {2010})}\BibitemShut {NoStop}%
\end{thebibliography}
\end{document}